\documentclass[prl,aps,twocolumn,superscriptaddress,floatfix]{revtex4}

\usepackage{color,graphicx,bm,amsmath}

\newcommand{\dsas}{\ensuremath{\Delta}}

\newcommand{\ld}{\ensuremath{L_{\mathrm{d}}}}

\begin{document}

\title{Critical currents and self organization in quantum Hall bilayers}

\author{P. R. Eastham}\affiliation{School of Physics, Trinity College, Dublin 2, Ireland.}

\author{N. R. Cooper}\affiliation{Cavendish Laboratory, 
University of Cambridge, Cambridge CB3 0HE, United Kingdom}

\author{D. K. K. Lee}\affiliation{Blackett Laboratory, Imperial
College London, London SW7 2AZ, United Kingdom}

\date{\today}

\begin{abstract} We present a theory of the critical interlayer
  tunneling current in a disordered quantum Hall bilayer at total
  filling factor one, allowing for the effect of static vortices. In
  agreement with recent experiments [Phys. Rev. B {\bf 80}, 165120
  (2009); \emph{ibid.}  {\bf 78}, 075302 (2008)], we find that this
  critical current is proportional to the sample area and is
  comparable in magnitude to observed values. This reflects the
  formation of a Bean critical state as a result of current injection
  at the boundary. We predict a crossover to a critical current
  proportional to the square-root of the area in smaller samples. We
  also predict a peak in the critical current as the electron density
  varies at fixed layer separation.
\end{abstract}

\maketitle

In a quantum Hall bilayer at total Landau level filling $\nu_T=1$,
Coulomb interactions induce a state with interlayer phase
coherence~\cite{murphy_many-body_1994,lay_anomalous_1994}, which can
be understood as a Bose-Einstein condensate of interlayer
excitons~\cite{fertig_energy_1989,eisenstein_bose-einstein_2004}. The
motion of excitons corresponds to counterflowing electrical currents
in the layers, so that excitonic supercurrents can give
dissipationless electrical transport. In particular, a finite
interlayer current $I$ at negligible interlayer voltage $V$ has been
predicted~\cite{wen_tunneling_1993,ezawa_lowest-landau-level_1993} and
recently confirmed by four-terminal
measurements~\cite{tiemann_critical_2008,tiemann_dominant_2009}. This
behavior, which is a form of Josephson
effect~\cite{wen_superfluidity_2003}, also appears as a zero-bias peak
in differential conductivity
measurements~\cite{spielman_resonantly_2000,eisenstein_evidence_2003}. The
low-voltage regime terminates when the interlayer current exceeds a
critical value $I_c$. Dissipation increases dramatically above $I_c$.

A key question has emerged over the nature of the critical current
$I_c$.  Recent experiments have shown that it is proportional to the
sample area~\cite{tiemann_dominant_2009,fincknote}. Within the
simplest model of a clean homogeneous bilayer, this scaling can only
be explained by assuming that the tunnel splitting is several orders
of magnitude smaller than expected~\cite{su_critical_2010}.  For
realistic splittings, tunneling is estimated to occur within a few
microns of the contact [Eq.~(\ref{eq:josephsonlength})] so that $I_c$
should not depend on the sample length in the direction of the
current~\cite{fil_josephson_2009,abolfath_critical_2003,su_critical_2010}.
Edge tunneling~\cite{rossi_interlayer_2005} has also been 
proposed, predicting linear scaling with the sample length but
not its area.

In this paper, we present a theory which produces a critical current
[Eq.  (\ref{eq:critcurrent})] that is proportional to the area of the
sample and of the correct order of magnitude, given reasonable
estimates for the parameters. A disorder-induced lengthscale, $\ld$,
emerges in our theory [Eq.~(\ref{eq:domainsize})]; this scale has no
counterpart in the clean system~\cite{su_critical_2010}. Our results
are also consistent with the observed dependence of $I_c$ on the
magnetic length and on an in-plane magnetic field.  A key test of our
theory is the prediction that $I_c$ should scale with the square-root
of the sample size for samples smaller than $\ld$.

The essential feature of our work is that we allow for static vortices
in the exciton superfluid, which will be nucleated by strong charge
disorder~\cite{eastham_vortex_2009,fertig_coherence_2005,
sternandbalents,fogler_josephson_2001}. Such
vortices play a crucial role for the critical current. They pin
 any injected supercurrents and sustain
dissipationless states, in much the same way that disorder pins
magnetic flux in
superconductors~\cite{tinkham_introduction_1996,larkin_pinning_1979,
vinokur_collective_1990},
or charge in charge-density
waves~\cite{fukuyama_dynamics_1978}. However, there
is a significant difference in the bilayer: the depinning force comes
from the injected charge current. Since these currents cannot
penetrate the bulk of the quantum Hall state, the depinning force is
applied \emph{only} at the sample boundary. Given this geometry, it is
a surprising feature of our results that the critical current $I_c$
can scale with the sample area.

\begin{figure}[t]
\includegraphics{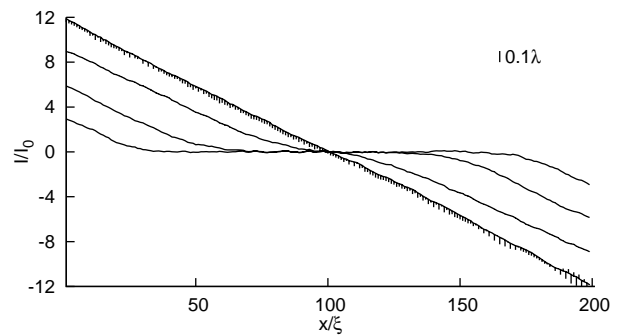}
\caption{Profile of counterflow current $I$ (lines) and interlayer voltage
  ($V\propto\dot\phi$, vertical bars, non-zero in top curve only) from 
  (\ref{eq:bilayerphased}) on a
  one-dimensional lattice with current injection at both ends, at a
  time $\approx 10^4/\lambda$ after the current is switched on. The
  injected counterflow currents at each boundary are $3, 6, 9, 12 I_0$
  for the four curves ($I_0=e\rho_s/\hbar\xi$). 
  The interlayer voltages vanish below
  a critical current $I_c$. Here, $9<I_c/I_0<12$. 
  $t\xi^2/\rho_s=0.4$, results averaged over 50
  realizations.\label{fig:1dmodel}}
\end{figure}

We will first present numerical results showing
that, in one dimension, currents injected at the boundary decay
linearly in space in the disordered state
(Fig.~\ref{fig:1dmodel}). This is analogous to the Bean critical state
in a superconductor. We argue that these special critical states are
generated by current injection from the boundary, and hence are selected
in the bilayer. Based on this numerical work,
we will then present a heuristic argument for the critical current, which we
generalize to two dimensions and compare with the experiments.

Our starting point is the energy functional 
\begin{equation}
H_{\rm eff}
  =\int \left[\frac{\rho_s}{2} (\nabla \phi)^2 - t \cos
    (\phi+\theta^0) \right]d^{D}\mathbf{r},
  \label{eq:phaseenergy_phi}
\end{equation} for the low-energy modes
of a bilayer containing pinned vortices. This form follows from the
clean model~\cite{wen_tunneling_1993} when the phase field of the
pinned vortices $\theta^0$ is subtracted out of the superfluid phase
$\theta$: $\phi=\theta-\theta^0$. The first term in
Eq. (\ref{eq:phaseenergy_phi}) is the superfluid stiffness while the
second describes the interlayer tunneling.  We assume that the vortex
field $\theta^0$ is disordered, with a correlation length
$\xi$ which we estimate to be of the order of the distance to the dopants, $d_d \approx\mathrm{200\ nm}$ in current
samples~\cite{eastham_vortex_2009}. The counterflow supercurrent
density above the ground state, $j_{\rm CF}$, and the interlayer
tunneling current density, $j_t$, are related to the phase field by:
\begin{equation}
j_{\rm CF}=\frac{e\rho_s}{\hbar} \nabla \phi\,,\qquad
j_t=et\sin (\phi+\theta^0)\,.
\label{eq:currents}
\end{equation}

A time-varying superfluid phase $\phi(t)$ gives rise to an interlayer
voltage difference $V$ \emph{via} the Josephson relation
$V=\hbar\dot{\phi}/e$.  Therefore, a state with a finite interlayer
current at zero interlayer voltage is time-independent, 
corresponding to a local minimum of the energy (\ref{eq:phaseenergy_phi}).
To investigate this possibility, we consider the dissipative
model \begin{equation} -\lambda \dot\phi=\frac{\delta
    H_{\mathrm{eff}}}{\delta \phi} = -\rho_s \nabla^2\phi+t \sin
  (\phi+\theta^0) ,
\label{eq:bilayerphased}
\end{equation}
whose stationary solutions $\dot \phi=V=0$ are the local minima
of Eq.~(\ref{eq:phaseenergy_phi}). The stationary equation
is the continuity equation stating
that the loss of counterflow current (first term) is accounted for by
interlayer tunneling  (second term).
The dissipative dynamics of Eq. (\ref{eq:bilayerphased}) is physically
a resistive shunt due to interlayer quasiparticle
tunneling. Additional dynamical terms are necessary in the finite
voltage
regime~\cite{fil_josephson_2009,jack_quantum_2005,sternandbalents,fogler_josephson_2001}. However,
it is the dissipation that determines the long-time limit at low
voltages, so that here we may use Eq.~(\ref{eq:bilayerphased}).

The boundary conditions for Eq. (\ref{eq:phaseenergy_phi}) come from
the current flows through the sample~\cite{su_critical_2010}. For
definiteness, we consider a tunneling geometry in which a current
$I_t$ is injected into the top layer at one corner and removed from
the bottom layer at the opposite corner. These current flows may be
written as superpositions of layer-symmetric and layer-antisymmetric
currents, 
\begin{equation*}
 I_{\mathrm{in(out)}}=\frac{I_t}{2}\left [ (1,1)\pm(1,-1)\right]
\end{equation*}
where the components
refer to the two layers. Thus, the tunneling experiment corresponds to
a flow of layer-symmetric current, with equal counterflow currents
$I=I_t/2$ injected by both the electron source and drain. In the
low-voltage regime, these counterflow currents will be the
supercurrents $j_{\rm CF}$ in Eq.~(\ref{eq:currents}). Since the
symmetric component cannot penetrate the bulk, the supercurrents are
injected at the boundary. As we shall see, the profile of their
injection along the boundary is unimportant.

The mechanisms leading to an extensive critical current for the
bilayer can be seen most clearly from a one-dimensional model. In
Fig.\ \ref{fig:1dmodel}, we show current and voltage profiles obtained
for the one-dimensional version of Eq.~(\ref{eq:bilayerphased}), on a
lattice of $L=200$ sites. These results are averaged over realizations
of the disorder $\theta_i^0$, which is taken independently on
different lattice sites. This corresponds to using the correlation
length $\xi$ of the continuum model as the lattice spacing.  The
natural unit of current is then $I_0=e\rho_s/\hbar\xi$.  For this
illustration, we take the tunneling strength $t\xi^2/\rho_s=0.4$. In
each realization, we start from an initial state in which the $\phi_i$
are random and independent, and equilibrate by integrating forwards in
time with the boundary conditions $\partial_x
\phi|_1=\partial_x\phi|_L=0$. To model the current injection in a
tunneling experiment, we then slowly increase the boundary conditions
to the final values
$\xi\partial_x\phi|_1=-\xi\partial_x\phi|_L=I/I_0$. For the lowest
three values of $I$ used, the dynamics reach a time-independent
solution, corresponding to the Josephson regime with vanishing
interlayer voltages. For too large $I$, these time-independent
solutions break down and the phase winds continuously at late
times. This corresponds to the breakdown of the d.c.~Josephson regime
and the appearance of a state with finite interlayer voltages.

The static states in Fig.\ \ref{fig:1dmodel} differ qualitatively from
those of the clean
model~\cite{fil_josephson_2009,bak_commensurate_1982},
$\theta^0=0$. In that case, there is a penetration depth
for the injected current of
\begin{equation}
\lambda_J\sim\sqrt{\rho_s/t}\,.
\label{eq:josephsonlength}
\end{equation}
Since the phase angle is periodic this implies a maximum injected
current density of $\partial_x\phi\sim \pi/\lambda_J$. In the
disordered case shown, however, the injected current decays linearly
close to the contacts, with a slope which is independent of the
injected current. Thus, an increase in the injected current is
accommodated by an increased current penetration into the sample. As
can be seen from the figure, this process continues until the currents
fill the entire sample. Beyond this point, further increases in
current cannot be accommodated by coherent tunneling and an interlayer
voltage develops. Since the current decays linearly with a constant
slope, the resulting $I_c$ in the one-dimensional model scales with
the sample length.

In the clean model, the breakdown of the stationary solutions can be
understood in terms of the injection of phase solitons at the
boundary~\cite{littlewood_metastability_1982,fil_josephson_2009}, which
propagate through the sample. Thus, the phase at any point varies in time, and the system develops an interlayer voltage by the a.c.~Josephson effect. However, for the disordered system, the solitons may be pinned by disorder. We now develop a heuristic theory of such pinning, which agrees with
our numerical work. 

We begin by recalling~\cite{fukuyama_dynamics_1978,imry_random-field_1975} the
form of the ground states of the random field XY model, Eq.~(\ref{eq:phaseenergy_phi}), in the weak disorder regime
$\xi\ll\lambda_J$ relevant for the bilayer. In this regime, the ground
state consists of ferromagnetic domains with polarized phase. The key
idea is that it is energetically costly to have phase twists at scales
shorter than the size, $\ld$, of these domains.  The energy cost for a
phase twist that varies over the scale $\ld$ is $\rho_s
\ld^{D-2}$ in $D$ dimensions. The typical tunneling energy of a polarized domain is
obtained by summing random energies in the range $\pm t\xi^D$ for its
$(\ld/\xi)^D$ correlation areas, giving $t\xi^D(\ld/\xi)^{D/2}$. Thus,
for $d<4$, the phase stiffness wins at short scales, and the domains
reach a finite size where the two energies balance. This gives the Imry-Ma scale:
\begin{equation} 
\ld\sim
\left(\frac{\rho_s}{t\xi^{D/2}}\right)^\frac{2}{4-D}=\left(\frac{\lambda_J^2}{\xi^{D/2}}\right)^{\frac{2}{4-D}}.
\label{eq:domainsize}
\end{equation} 
In this ground state of polarized domains, the average coarse-grained phase over a domain is chosen such that the tunneling energy $H_t$ of each domain is minimized.
Since $\delta H_t/\delta\phi(\mathbf{r})$ is the tunneling current at position
$\mathbf{r}$, the total tunneling current over the domain vanishes. 

We now consider how the system changes in the presence of an injected
current. The injected counterflow will cause the phase to twist away
from its equilibrium value, leading to finite tunneling currents.  We
assume that the configuration remains smooth on the scale $\ld$, and
hence average Eq.~(\ref{eq:bilayerphased}) over each domain.  The
second term in Eq.~(\ref{eq:bilayerphased}) becomes
$(t/\ld^D)(\ld/\xi)^{D/2} f(\bar\phi)$ where $\bar\phi$ is the
deviation of the coarse-grained phase from its equilibrium value, and
the range of $f(\bar\phi)$ is typically $[-1,1]$.
For a dissipationless state ($\dot\phi=0$),
the coarse-grained phase $\bar\phi$ should therefore obey:
\begin{equation}
-\ld^2\nabla^2\bar\phi+f(\bar\phi) = 0\;.
\label{eq:staticf} 
\end{equation}

The source term in Eq. (\ref{eq:staticf}) describes the loss of
injected current due to tunneling in a domain.  As discussed above,
current injection induces counterflow currents and hence phase twists.
Since it is energetically costly to introduce phase twists in a
domain, the domain at the boundary will respond by rotating uniformly,
increasing its tunneling current, thereby reducing the counterflow
current. This process continues until the tunneling in the domain
saturates, so that $|f|\sim 1$. The residual counterflow currents will
be transmitted further into the sample, causing the domains there to
rotate in a similar way. Thus, we argue that forcing at a boundary
leads to a self-organized critical state, in which the driven part of
the system sits at the threshold $|f|\sim 1$.

In one dimension, this argument means that Eq.~(\ref{eq:staticf})
would give an average counterflow current $\rho_s\nabla\bar\phi$ that
decreases linearly from the boundary. To be more precise, it predicts
a linear $I/I_0$ in the saturated regions, with a slope
$-\xi^{-1}(t\xi^2/\rho_s)^{4/3}$. This is qualitatively consistent
with the numerical results shown in Fig.~\ref{fig:1dmodel}.  Note
that, in one dimension, Eq.~(\ref{eq:staticf}) describes a harmonic
chain with random static
friction~\cite{coppersmith_phase_1990}. The
process described above is simply the transmission of forces when such
a chain is pushed at its ends.

As we now describe, the generalization of this argument to two dimensions
will account for the critical current of the bilayer, as measured by
Tiemann \emph{et al}.~\cite{tiemann_dominant_2009}. (We see similar
behavior in two-dimensional simulations, albeit with large disorder
fluctuations.) In our scenario of saturated domains, the static state
only breaks down when the final domain in the sample exceeds
threshold. Therefore, we may determine the critical current by setting
$f=1$ everywhere in Eq. (\ref{eq:staticf}). Integrating over space, we
see that the critical current, defined as the total injected current
at threshold, functionly counts the number of domains in the sample,
and is independent of the precise geometry. We find, for its order of
magnitude,
\begin{equation} 
I_c\sim \frac{e \rho_s}{\hbar}  \frac{S}{\ld^D}, 
\label{eq:critcurrent} 
\end{equation}
where $S$ is the sample area in 2D and sample length in 1D. Note that
this is a natural form for an extensive critical current, composed of
the system size, the characteristic length \ld, and the microscopic
current scale $e\rho_s/\hbar.$ The same form can be seen in Eq. (13)
of Ref.  \onlinecite{fogler_josephson_2001}.  However, it is not clear
whether the result there applies to a bilayer driven at its boundary.

For Eq. (\ref{eq:critcurrent}) to apply, the sample should
comprise many domains. If a dimension $L_x$ is smaller than the domain
size $\ld$ then the second term in Eq. (\ref{eq:staticf}) should be
multiplied by $\sqrt{\ld/L_x}$, because the total tunneling current of
the domain is cut off at the sample width $L_x$. This gives 
\begin{equation}
  I_c\sim \frac{e \rho_s}{\hbar}
  \sqrt{\frac{L_x}{\ld}}\frac{L_y}{\ld}\quad
  (\mbox{quasi-1D:\ } L_x\ll \ld \ll L_y)\,.
  \label{eq:critcurrentq1d}
\end{equation} 
Similarly, for a sample containing only a single domain, 
\begin{equation} 
I_c\sim
  \frac{e \rho_s}{\hbar} \sqrt{\frac{L_x L_y}{\ld^2}}
  \quad(\mbox{for\ } L_x, L_y \ll \ld).
\label{eq:critcurrentq0d}
\end{equation}

To compare Eq.~(\ref{eq:critcurrent}) with the experiments, we start
from the microscopic theory~\cite{joglekar_microscopic_2001} for the
zero-temperature values of the stiffness, $\rho^0_s$, and order
parameter, $m_x = \langle\cos\phi\rangle$, for a homogeneous
bilayer. The latter renormalizes the tunneling strength $t^0$ so that
$t^0 = \dsas_0 m_x/2\pi l^2$ where $\dsas_0$ is the single-particle
tunnel splitting. This theory does not predict accurately the critical
layer separation for the loss of interlayer coherence but should be
reasonable away from the critical point.  Owing to the strong charge
disorder, we expect the incompressible quantum Hall phase to occupy a
small fraction of the sample, with the remainder occupied by puddles
of compressible electron
liquid~\cite{eastham_vortex_2009,fertig_coherence_2005}.  We suppose
that the incompressible phase forms a
network~\cite{fertig_coherence_2005} of channels of size $l$
separating puddles of size $\xi \approx d_d$, the distance to the
dopants.  Thus, the effective parameters in the energy
[Eq.~(\ref{eq:phaseenergy_phi})] for the disordered bilayer should be
$t \sim (l/d_d)t^0$ and $\rho_s\sim (l/d_d) \rho^0_s$ ($\approx 20
\mathrm{mK}$ at a layer separation $d\approx l$).  The microscopic
current scale is then $e\rho_s/\hbar\approx 0.5\; \mathrm{nA}$.

\begin{figure}[hbt]
\includegraphics{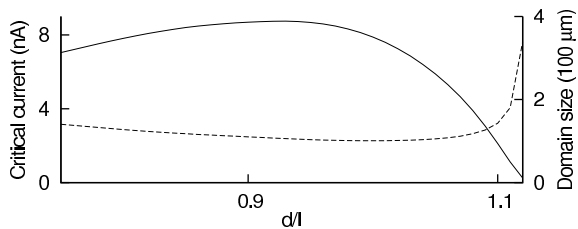}
\caption{Predicted critical current (solid, left axis) and domain size
  (dotted, right axis), for area $S=0.1\, \mathrm{mm^2}$,
  $d=28\,\mathrm{nm}$, $\xi\, (\approx d_d) = 200\,\mathrm{nm}$, and
  bare tunneling $\dsas_0 =150\,\mu\mathrm{K}$. The stiffness and
  tunneling renormalization are taken from theory
  \cite{joglekar_microscopic_2001}, and scaled by the network
  geometrical factor (see text).  }
\label{fig:critcurr}
\end{figure}

Figure~\ref{fig:critcurr} shows our estimates for the domain size
$\ld$ and the critical current $I_c$, as functions of the ratio of
interlayer separation to magnetic length, $d/l$. We see that the
domains ($\ld^2 \lesssim 0.01$\;mm$^2$) are indeed not larger than the
samples areas of $(0.01 - 1)$\;mm$^2$ so that the results are
consistent with area scaling for $I_c$.  Moreover, $I_c$ has the
correct order of magnitude compared to the observed
values~\cite{tiemann_dominant_2009} of $0.1-10\;\mathrm{nA}$.

Interestingly, $I_c$ has a peak as a function of $d/l$ which is also
suggested in the experimental data~\cite{tiemann_dominant_2009}. This
feature appears robust: it arises from the increase in $L_d$ as $d/l$
is reduced, caused by the increase in $\rho_s$. However, the
peak position depends on the precise variation of model parameters
with $d/l$. For example, the variation in $\rho_s$ may cause some
variation in vortex density~\cite{eastham_vortex_2009} and hence
$\xi$, pushing the peak to smaller values of $d/l$.

Our theory also has implications for the effect of an in-plane magnetic
field $B_\parallel$. This introduces a lengthscale $l_\parallel=
hc/eB_\parallel d$, which is the length of a loop enclosing a
flux quantum in the cross-section of a bilayer.
Within our theory, the system should be insensitive to $B_\parallel$ 
unless $l_\parallel<\xi$. At such fields, circulating
tunneling currents are set up between the two layers, so that the
\emph{net} tunneling current in a domain is reduced. Therefore, 
coherent tunneling should be suppressed for $B_\parallel >
hc/ed\xi \sim$ 0.7\;T, which is consistent with
experiments~\cite{spielman_observation_2001} where the enhanced
tunneling decreases above 0.5\;T.

In conclusion, we have presented a theory of the critical interlayer
currents in a disordered quantum Hall bilayer with static pinned
vortices. We find that, because the current is injected at the
boundary, coherent tunneling saturates in the current-carrying region,
leading to a Bean critical state. This results in an extensive
critical current for sufficiently large samples (in contrast to the
clean limit~\cite{su_critical_2010} where area scaling holds for small
samples). The magnitude of the critical current is consistent with
experiments. We predict that area scaling does not hold when the
samples become smaller than the phase-pinned domains, and also that
the critical current peaks in the interlayer coherent phase.

We thank P.~B.~Littlewood for helpful discussions. This work was
supported by EPSRC-GB (EP/C546814/01) and Science Foundation
Ireland (SFI/09/SIRG/I1952).

\vspace{-0.25in}\bibliography{qhcritj_condmat}
\end{document}